\documentclass[st,twocolumn]{jpsj3}
\usepackage{txfonts}
\usepackage{bm}
\usepackage{graphicx}
\usepackage{pdfpages}

\newcommand{\calD}{\mathcal{D}}
\newcommand{\calI}{\mathcal{I}}
\newcommand{\calN}{\mathcal{N}}
\newcommand{\bfunc}{{\bm f}}
\newcommand{\bs}{{\bm s}}
\newcommand{\bx}{{\bm x}}
\newcommand{\bw}{{\bm w}}
\newcommand{\bxi}{{\bm \xi}}
\newcommand{\bnabla}{{\bm \nabla}}
\newcommand{\btheta}{{\bm \theta}}
\newcommand{\rmd}{\mathrm{d}}
\newcommand{\rme}{\mathrm{e}}
\newcommand{\rmi}{\mathrm{i}}
\newcommand{\RE}{\mathrm{Re}}
\newcommand{\IM}{\mathrm{Im}}
\newcommand{\ar}{a_{\mathrm{R}}}
\newcommand{\ai}{a_{\mathrm{I}}}

\title{Stochastic quantization and diffusion models}

\author{Kenji Fukushima and Syo Kamata}
\inst{
Department of Physics, The University of Tokyo\\
7-3-1 Hongo, Bunkyo-ku, Tokyo 113-0033, Japan
}

\abst{
  This is a pedagogical review for the possible connection between the stochastic quantization in physics and the diffusion models in machine learning.
  For machine-learning applications, the denoising diffusion model has been established as a successful technique, which is formulated in terms of the stochastic differential equation (SDE).
  In this review, we focus on an SDE approach used in the score-based generative modeling.  Interestingly, the evolution of the probability distribution is equivalently described by a particular class of SDEs, and in a particular limit, the stochastic noises can be eliminated.
  Then, we turn to a similar mathematical formulation in quantum physics, that is, the stochastic quantization.  We make a brief overview on the stochastic quantization using a simple toy model of the one-dimensional integration.  The analogy between the diffusion model and the stochastic quantization is clearly seen in this concrete example.
  Finally, we discuss how the sign problem arises in the toy model with complex parameters.  The origin of the difficulty is understood based on the Lefschetz thimble analysis.  We point out that the SDE is not invariant under the variable change which induces a kernel and a special choice of the kernel guided by the Lefschetz thimble analysis can reduce the sign problem.
}

\begin{document}
\maketitle

%%%%%%%%%%
\section{Introduction}

Generative modeling is widely used for practical applications, among
which denoising diffusion probabilistic models
(DDPMs)~\cite{ho2020denoising} based on physical processes of
non-equilibrium dynamics~\cite{sohl2015deep} appear to be a
physics-friendly  formulation with the Langevin equation or the
stochastic differential equation (SDE).  In particular, the approach
with score
matching~\cite{song2019generative,ramzi2020denoising,song2020score}
further extends a physics intuition for denoising processes.  The
interesting observation is that the reverse of the stochastic noising
process also follows the SDE once the score matching is achieved.  For a
proof of the reverse process in the language of physics, see a recent
work~\cite{Hirono:2024zyg}.

The denoising process allows for sampling of generated data.  It is
shown~\cite{song2020denoising} that a wider class of diffusion
processes can lead to the equivalent distribution of sampling data,
which even include an ordinary differential equation (ODE) that is
fully deterministic.  Clearly, such a deterministic formulation of
denoising diffusion implicit models (DDIMs)~\cite{song2020denoising}
can perform sampling much faster.  To understand the equivalence, as
we will review later, the crucial point is that the probability of
samples governed by the Langevin equation should evolve with the
Fokker-Planck equation.  Therefore, if the same Fokker-Planck equation
is derived from a class of SDEs with free parameters, in principle, the generated data should exhibit the same quality.

In this review, we pay attention to an analogy between the diffusion model and the stochastic quantization in physics~\cite{Damgaard:1987rr} and discuss a potential interplay.  It should be noted that the idea to accelerate configuration generation in lattice field theory has been tested within the generative diffusion model~\cite{Wang:2023exq}.  Then, it has been clearly recognized that the diffusion model can be interpreted as the probability evolution in the stochastic quantization.

In physics, the quantum effect is often called ``fluctuation'' and in the quantization procedure such quantum fluctuations are integrated out.  Then, one may naturally be tempted to associate a Brownian-type motion with the quantum effect.  The reformulation of the Schr\"{o}dinger
equation in terms of classical noises is dated back to a seminal
work~\cite{PhysRev.150.1079} more than half a century ago.  In
general, however, it is impossible to map all the quantum effect to
classical fluctuations unless an extra dimension along the quantum axis
is introduced, which underlies a general idea of holographic
principle.  Such a duality between a quantum theory in $d$ dimensions
and a classical theory in $(d+1)$ dimensions is not limited to the
gauge-gravity correspondence, but many useful examples can be found in
various contexts, such as the renormalization group flow.  The
stochastic quantization belongs to this category of machinery to
quantize theories with classical variables with the quantum axis in an extra dimension.

The key equation in the stochastic quantization is the SDE and the mathematical structures are quite similar to the DDPMs.  Thus, it is an intriguing direction of research to think about the mutual interplay between two approaches. Such an attempt has been just launched recently~\cite{Wang:2023exq}, and this article is expected to serve as
a starting point for further productive interactions of two formulations in different communities.

%%%%%%%%%%
\section{A Review of Denoising Diffusion Probabilistic Models}

We make a brief review of the DDPMs, especially the score-matching
modeling according to the standard literature~\cite{song2020score}.
The forward noising diffusion is described by the following Langevin
equation or the SDE:
\begin{equation}
  \rmd{\bx}_t = \bfunc (\bx_t, t)\rmd t + g(t)\, \rmd\bw_t\,.
  \label{eq:SDEw}
\end{equation}
Here, $\bfunc(\bx_t, t)$ represents the drift term and $g(t)$ is the
diffusion coefficient.  In general $g(t)$ can take a matrix structure
but we treat it as a one-component function for simplicity.  The last
term involves a stochastic variable $\rmd\bw_t$ that is the Wiener
process.  Importantly,
both $\bfunc(\bx_t, t)$ and $g(t)$ are time-dependent so that the
distribution of $\bx_t$ converges to the normal distribution after
all.  In the physics notation, a more familiar form of
Eq.~\eqref{eq:SDEw} would be
\begin{equation}
  \dot{\bx}_t = \bfunc (\bx_t, t) + g(t)\, \bxi_t\,.
  \label{eq:SDE}
\end{equation}
Here, $\rmd\bw_t = \bxi_t \rmd t$ and $\rmd\bw_t^2=\rmd t$
symbolically in the It\^{o} calculus, and $\dot{\bx}_t$ represents the
time-derivative of $\bx_t$.  We note that $t$ will turn out to be a
fictitious time in later discussions about the stochastic quantization.

It is convenient to establish a relation between the SDE and the
evolution equation for the probability distribution that $\bx_t$
follows.  The evolution equation is known as the
Fokker-Planck-Kolmogorov (FPK) equation, which is commonly called just
the Fokker-Planck equation in physics or the forward Kolmogorov
equation in stochastic theory.  The explanation we summarize below is
based on the argument in the review~\cite{Damgaard:1987rr} in which
the Stratonovich calculus is assume to use the ordinary chain rule.
To see the correspondence to the Fokker-Planck equation, we consider the average of some function
$A(\bx_t)$ with respect to the noise $\bxi_t$ as denoted in physics by
\begin{equation}
  \langle A(\bx_t)\rangle := \int D\bxi\, A(\bx_t)\,
  \exp\biggl( -\frac{1}{2}\int ||\bxi_t||^2 \rmd t \biggr)\,,
  \label{eq:noise_average}
\end{equation}
where $\bx_t$ is a solution of Eq.~\eqref{eq:SDE} and thus $\bxi_t$
dependence is implicitly implemented through $\bx_t$.
Using this, we can also introduce the probability $p_t(\bx)$ from
\begin{equation}
  \langle A(\bx_t)\rangle =: \int \rmd\bx\, A(\bx) p_t(\bx)\,.
\end{equation}
The right-hand size of the above is often denoted as
$\mathbb{E}_{p_t}[A(\bx)]$.  We note that the $t$-dependence is
incorporated only in $p_t(\cdot)$ and $\bx$ is just the integration
variable.  Now, our task is to find an equation to describe the
time-evolution of $p_t(\bx)$ that is consistent with
Eq.~\eqref{eq:SDE}.  The time derivative of $\langle A(\bx_t)\rangle$
is immediately given by
\begin{align}
  \int \rmd \bx\, A(\bx)\, \dot{p}_t(\bx)  
  &= \biggl\langle \frac{\partial A(\bx_t)}{\partial \bx_t}
  \cdot \dot{\bx}_t \biggr\rangle \notag\\
  &= \biggl\langle \frac{\partial A(\bx_t)}{\partial \bx_t}
  \cdot \bigl(\bfunc(\bx_t,t) + g(t)\, \bxi_t\bigr) \biggr\rangle \,.  
\end{align}  
From the definition~\eqref{eq:noise_average}, we can see:
\begin{equation}
  \biggl\langle \frac{\partial A(\bx_t)}{\partial \bx_t}
  \cdot \bxi_t \biggr\rangle
  = \biggl\langle \frac{\partial}{\partial\bxi_t} \cdot
  \frac{\partial A(\bx_t)}{\partial \bx_t}
  \biggr\rangle
  = \biggl\langle \frac{\partial \bx_t}{\partial\bxi_t}
  \frac{\partial^2 A(\bx_t)}{\partial \bx_t^2}
  \biggr\rangle \,. 
\end{equation}
Here, we used the integration by part so that $\bxi_t$ comes down from
the exponential function.  Using
$\partial \bx_t/\partial \bxi_t=g(t)/2$, where $1/2$ appears from
$\theta(0)=1/2$ with the Heavisite step function, we arrive at
\begin{align}
  & \int \rmd \bx\, A(\bx)\, \dot{p}_t(\bx)  \notag\\
  &= \int \rmd \bx\, A(\bx) \frac{\partial}{\partial \bx}
    \cdot \biggl( -\bfunc(\bx, t) + \frac{g(t)^2}{2}
    \frac{\partial}{\partial\bx} \biggr) \,p_t(\bx)\,.
\end{align}
The above is a heuristic argument for physicists, while a more secure
derivation utilizes the It\^{o}'s formula.  In any case, because this
expression should hold for any function $A(\bx)$, we conclude the
following Fokker-Planck equation:
\begin{equation}
  \dot{p}_t(\bx) = -\bnabla\cdot \biggl[ \bfunc(\bx, t)\, p_t(\bx)
  - \frac{g(t)^2}{2} \bnabla p_t(\bx) \biggr]\,.
  \label{eq:FP}
\end{equation}
Here, we use $\bnabla$ instead of $\partial/\partial\bx$ for notational
brevity.  It is clear that the Langevin equation~\eqref{eq:SDE} and
the Fokker-Planck equation~\eqref{eq:FP} have equivalent physical
contents.  Nevertheless, solving the Fokker-Planck equation
numerically demands huge computational costs, and the Langevin
equation is more tractable.  We comment that the Fokker-Planck
equation appears in a wide range of physical systems including even
the high-energy scattering process of quarks and
gluons~\cite{Kovchegov:2012mbw}, for which the Fokker-Planck equation
in non-Abelian group space called the JIMWLK equation is an
established theoretical tool.  It is impossibly difficult to solve
such a complicated functional equation, but the equivalent rewriting
to the Langevin equation paves a path for feasible numerical
simulations~\cite{Blaizot:2002np,Rummukainen:2003ns}.  We leave this
comment with a hope that the generative modeling could be useful even
for such resummation programs in high-energy small-x physics.

The Fokker-Planck equation tells us a condition for the drift and the
noise terms.  If we na\"{i}vely disturb $\bx_t$ with noise, the
distribution of $\bx_t$ may simply spread widely.  To make the reverse
process well organized, we require an asymptotic form at $t\to \infty$
to take the normal distribution, i.e.,
$p_{t\to\infty}(\bx) \propto e^{-||\bx||^2/(2\sigma^2)}$.  We can plug
this form of the normal distribution into the Fokker-Planck equation,
and then we can deduce the condition;  we should choose
$\bfunc(\bx,t)$ and $g(t)$ such that
$\lim_{t\to\infty} \bfunc(\bx,t)/g(t)^2 = -\bx/(2\sigma^2)$.  For the
normal distribution with $\sigma^2=1$, the simple choice of
time-dependent coefficients, i.e., the SDE schedulings is:
\begin{equation}
  \bfunc(\bx,t) = f(t)\,\bx = -\frac{\beta}{2}t\,\bx\,,\qquad
  g(t) = \sqrt{\beta t}\,.
\end{equation}
Technically, it is notable that the drift term, $\bfunc(\bx,t)$, is a
linear function of $\bx$, i.e., it is \textit{affine}.  This choice
enables us to derive some useful analytical formulas.  We can
arbitrarily take $\beta$ which just controls the scale of time
evolution.  In this work, we fix $\beta=20$ following the
convention~\cite{Hirono:2024zyg}.  The important feature of the affine
drift term is that the mean, ${\bm m}(t)$, and the variance,
$\sigma(t)^2$, of the conditional probability, $p_{t|0}(\bx|\bx_0)$,
is solved.  From the Fokker-Planck equation~\eqref{eq:FP}, it is
straightforward to derive the following differential equations:
\begin{align}
  \dot{\bm m} &= \mathbb{E}[\bfunc(\bx,t)]
  = f(t)\,{\bm m}\,, \\
  \dot{\sigma}^2 &= 2\mathbb{E}[\bfunc(\bx,t)\cdot (\bx-{\bm m})] + g(t)^2
  = 2f(t)\, \sigma^2 + g(t)^2\,.
\end{align}
The solution of above two equations is found to be
\begin{equation}
  {\bm m} = \alpha(t)\,\bx_0\,,\qquad
%  = \rme^{\int_0^t f(\xi) \rmd \xi}\, \bx_0\,,\quad 
  \sigma(t)^2 = \alpha(t)^2 \int_0^t \frac{g(\xi)^2}{\alpha(\xi)^2}
  \rmd\xi\,,
  \label{eq:sol_affine}
\end{equation}
where
\begin{equation}
  \alpha(t) = \rme^{\int_0^t f(\xi) \rmd \xi}\,.
\end{equation}
These expressions will appear in the loss function in what follows
below.

From now, we shall continue the explanation by taking a concrete
example.  From the analogy to the stochastic quantization as we
discuss later, we set up the simplest one-dimensional problem such
that the probability distribution is given by a double-well potential
form, i.e.,
\begin{equation}
  p_0(x) = Z^{-1}(a,b)\, \rme^{-S(x;a,b)}\,, \qquad
  S(x;a,b) = a x^2 + b x^4\,.
  \label{eq:p0}
\end{equation}
We assume $a, b \in \mathbb{R}$ for the moment, and when we analyze
the sign problem later, we will generalize them to complex numbers.
The normalization is $Z(a,b)=\int \rmd x\, \rme^{-S(x;a,b)}$ which
converges for $b >0$ or $a > 0$ if $b=0$.  This
integral is expressed in terms of special functions, if $a>0$
and $b>0$, as
\begin{equation}
  Z(a,b) = \frac{1}{2} \sqrt{\frac{a}{b}} \rme^z
  K_{1/4} (z)
  \label{eq:Zsol}
\end{equation}
with $z=a^2/(8b)$.  Interestingly, the above expression holds for
complex $a$ and $b$ as long as $\RE{a}>0$ and $\RE{b}>0$.

In this study we rather consider a bit more nontrivial case with
$a<0$.  To make our analyses concrete, we specifically choose
$a=-4$ and $b=1$ to visualize the symmetry-breaking-type potential.

In this toy model the true distribution $p_0(x)$ is already known in
Eq.~\eqref{eq:p0}, and let us explain the step-by-step procedures
in the score-based diffusion model.

\begin{figure}
  \centering 
  \includegraphics[width=0.9\columnwidth]{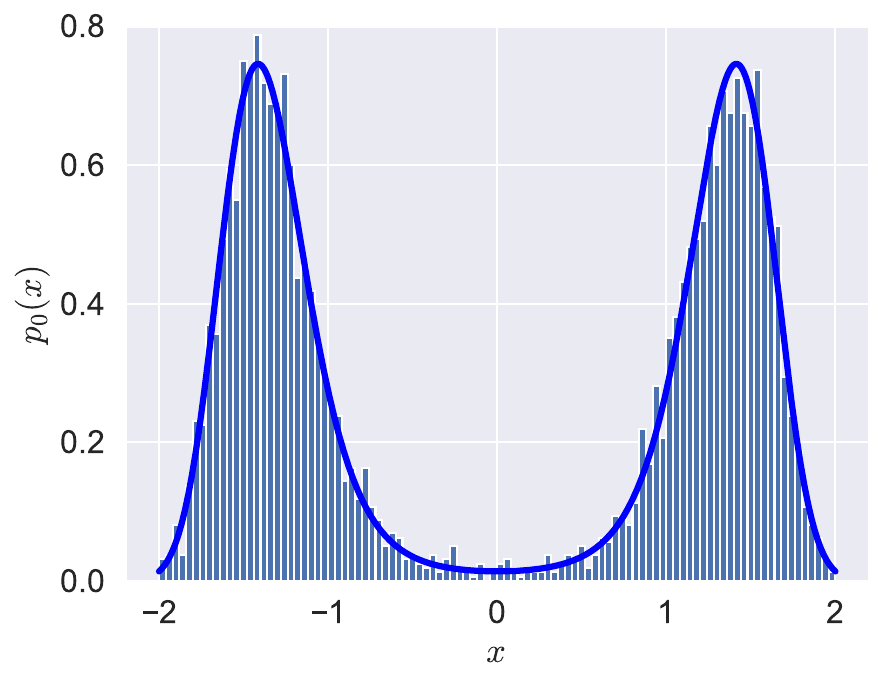}
  \caption{Randomly sampled 4000 points according to the given 
    $p_0(x)$ with $a=-4$ and $b=1$ that is shown by the solid line.
    The  histogram is plotted from $x=-2$ to $+2$ with the bin size
    equally divided by $100$ intervals.}
  \label{fig:data}
\end{figure}

First, we sample points from $p_0(x)$, which corresponds to sampling
image data for the training.  Although we already have $p_0(x)$ in the
present exercise, we usually do not know what $p_0(x)$ looks like.  We can
collect the data for the training purpose, assuming the existence of
some probability distribution that rules the data.  At first glance,
it may sound like an unmanageable task, but amazingly, the model can be
trained efficiently with the given training data.  This is a vital feature for the
practical usage.  In Fig.~\ref{fig:data}, we show the example of random
sampling according to $x \sim p_0$.

% --- figure ---%
% --- figure ---%

These sampled points, $\{ x_i \}$, give the initial values;
$x_{i, t=0}=x_i$, and the SDE leads to $x_{i, t}$ for later time.  We
shall integrate the SDE up to $t=T$.  For sufficiently large $T$, it
is expected that $x_{i, T} \sim \mathcal{N}(0, 1)$;  remember that
$f(\bx,t)$ and $g(t)$ were chosen in such a way.  The non-trivial
question is how $p_t(x)$ should behave in the intermediate time
region.  In principle, the Fokker-Planck equation~\eqref{eq:FP}
uniquely solves $p_t(x)$ for a given initial condition.  However, it
is generally difficult to solve $p_t(x)$ as it is, and we can
translate the problem into the optimization problem of the score
function,
\begin{equation}
  \bs_t(\bx; \btheta) \approx \bnabla \ln p_t(\bx)\,,
\end{equation}
where $\btheta$ denotes fitting parameters to approximate $p_t(\bx)$.
Thus, in the DDPM, the loss function is the deviation of $\bs_t$ from
$\bnabla\ln p_t$.  The $L^2$-norm yields the loss function as follows:
\begin{equation}
  L(\btheta) = \mathbb{E}_{p_t} \Bigl[ || \bs_t(\bx;\btheta)  
  - \bnabla \ln p_t(\bx) ||^2 \Bigr]\,.
  \label{eq:DDMloss}
\end{equation}
In the case of the Implicit Score Matching (ISM), we can evaluate this
loss function only with $\{\bx_i\}$ through the following rewriting:
\begin{align}
  L_{\mathrm{ISM}}(\btheta)  
  &= \int \rmd\bx_t\, p_t(\bx) \Bigl( \bs_t^2 - 2\bs_t\cdot \frac{\bnabla  
    p_t}{p_t} \Bigr) + \text{(const.)} \notag\\
  &= 2 \mathbb{E}_{p_t} \Bigl[ || \tfrac{1}{2}\bs_t^2 + \bnabla\cdot \bs_t  
    ||^2 \Bigr] + \text{(const.)}
  \label{eq:Lism}
\end{align}
Here, $\text{(const.)}$ represents the terms involving not $\bs_t$ but
$p_t$ only which are independent of $\btheta$.  From the first to the
second line, the integration by part is performed to move $\bnabla$
onto $\bs_t$.  In this final form, remarkably, there is no explicit
$p_t(\bx)$ in the function itself but it appears in the weight.  Thus,
this expectation value is well approximated by the training data
$\{\bx_i\}$ under the assumption that the training data obey
$p_t(\bx)$.  On the other hand, in the method referred to as the
Denoising Score Matching (DSM), the loss function is given by
\begin{align}
  L_{\mathrm{DSM}}(\btheta) 
  &= \frac{1}{N}\sum_{i=1}^N || \bs_t(\bx_{i,t};\btheta) 
    - \bnabla \ln p_{t|0}(\bx_{i,t}|\bx_{i,0}) || \notag\\
  &= \frac{1}{N}\sum_{i=1}^N \Bigl|\Bigl| \bs_t(\bx_{i,t};\btheta)
    + \frac{\bx_{i,t} - \alpha(t) \bx_{i,0}}{\sigma(t)^2} \Bigr|\Bigr|\,,
    \label{eq:Ldsm}
\end{align}
where the explicit solution of $p_{t|0}(\bx_{i,t}|\bx_{i,0})$ is used
with ${\bm m}$ and $\sigma(t)^2$ in Eq.~\eqref{eq:sol_affine}.  This
is again trainable with $\{x_{i,t}\}$.
This form~\eqref{eq:Ldsm} is more advantageous than Eq.~\eqref{eq:Lism} because of the lack of the gradient.

\begin{figure}
  \centering 
  \includegraphics[width=0.9\columnwidth]{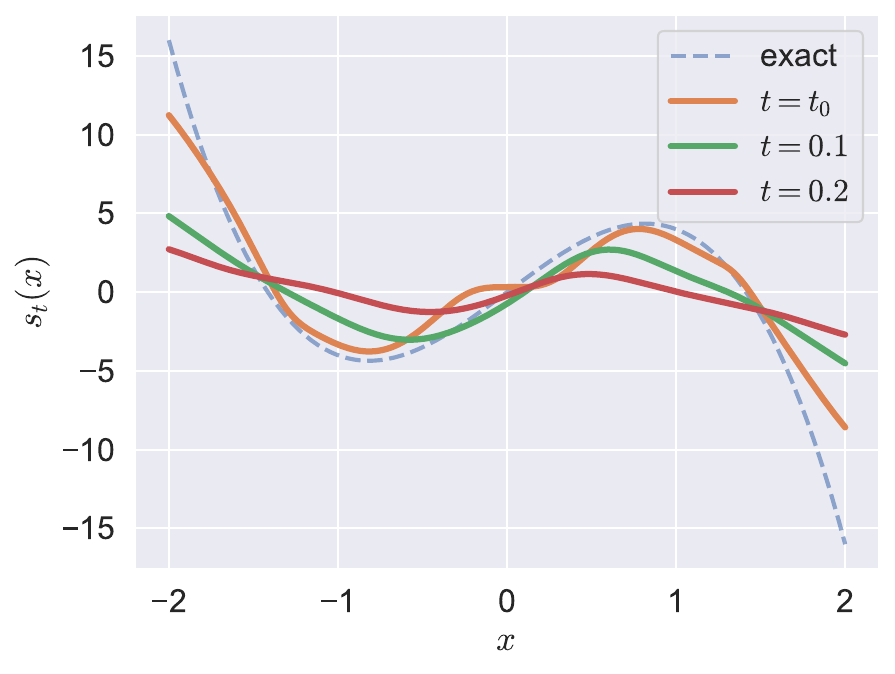}
  \caption{Trained score functions at $t=t_0$, $0.1$, $0.2$,
    respectively, with the exact answer,
    $\partial_x \ln p_0(x) = -2ax-4bx^3$, overlaid by the dashed
    line.}
  \label{fig:score}
\end{figure}

The python code for two-dimensional numerical simulations in the preceding
work~\cite{Hirono:2024zyg} is provided in public, and we adapted the
code for our one-dimensional simulations.  We employ a neural network
to represent $s_t(x)$ as defined in the original code.  Specifically,
in the previous literature~\cite{Hirono:2024zyg}, the choice of the
neural network is
$[x,t]\to (\texttt{Dense}(128)\to \texttt{Swish})^3\to \texttt{Dense}(1)\to s_t(x)$.
Here, $\texttt{Swish}$ is a function also called SiLU given by
$x/(1+\rme^{-x})$ that looks like ReLU but has no vanishing gradient.
Since our current problem is one-dimensional, one might think that far
simpler neural networks could work fine, but we numerically found that
the performance was not satisfactory if we reduced the layer size.  We
then train the model from $t=t_0=0.01$ to $t=T=1$ with
$\Delta t=(T-t_0)/N_t$ where $N_t=10^3$.  We choose the loss function in Eq.~\eqref{eq:Ldsm} and take the batch size as $N=32$.  For the training, the epoch number is $300$.  In our present
problem, $p_0(x)$ is set by hand, and the corresponding
one-dimensional score is $\partial_x \ln p_0(x) = -2ax-4bx^3$.
Figure~\ref{fig:score} shows the trained score functions at
$t=t_0$, $0.1$, $0.2$, respectively, together with the exact answer.
We can confirm that $s_0(x)$ at $t=t_0$ well approximates
$\partial_x \ln p_0(x)$.  From the imposed condition,
$\lim_{t\to\infty} \bfunc(\bx,t)/g(t)^2=-\bx/(2\sigma^2)$, we see that
$\bs_{t\to\infty}(\bx)=-\bx$ asymptotically.  Actually, in
Fig.~\ref{fig:score}, the score function $s_t(x)$ at $t=0.2$ is
already close to this asymptotic behavior of $s_{t\to\infty}(x)=-x$.

% --- figure ---%
% --- figure ---%

\begin{figure}
  \centering
  \includegraphics[width=0.8\columnwidth]{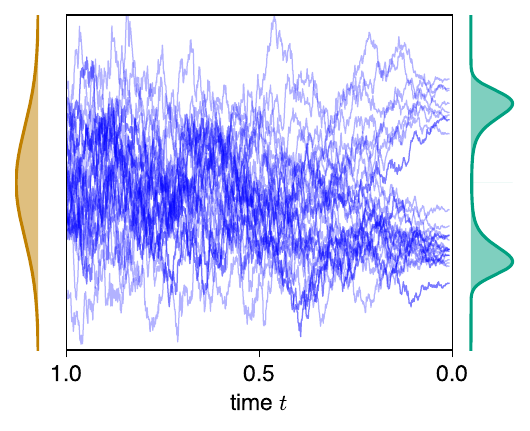}
  \caption{Randomly sampled $30$ trajectories for the reverse process  
    to recover the original probability distribution (depicted by the  
    green shaded region) from the normal distribution (depicted by the  
    brown shaded region).}
  \label{fig:reverse}
\end{figure}

Once $s_t(x)$ is trained well, the denoising or the reverse process is
described by the following SDE:
\begin{equation}
  \dot{\bx}_t = \bfunc(\bx_t, t) - g(t)^2 \bs_t(\bx_t; \btheta)  
  + g(t)\, \bar{\bxi}_t\,,
  \label{eq:SDErev}
\end{equation} 
where $t$ decreases from $t=T$ to $t=t_0$.  Figure~\ref{fig:reverse}
shows randomly sampled $30$ trajectories with initial
$x_{i, T} \sim \calN(0,1)$.

% --- figure ---%
% --- figure ---%

The physical meaning is transparent;  $\ln p_t$ is regarded as a
negative potential energy and $\partial_x \ln p_t$ is thus a force.
Therefore, in the $x$-region with $s_t(x)>0$, the particle at
$x_{i,t}$ tends to move to the positive-$x$ direction, while in the
$x$-region with $s_t(x)<0$, the tendency is opposite.  As a
consequence, the distribution becomes denser near $x_0$ with
$s_t(x_0)\approx 0$ when $\partial_x s_t(x_0) < 0$, and the
distribution is diluted around $x_0$ with $s_t(x_0)\approx 0$ when
$\partial_x s_t(x_0) > 0$.  In this way, one can easily associate
$s_t(x)$ in Fig.~\ref{fig:score} with the two-peak structure of
$p_0(x)$ in Fig.~\ref{fig:data}.

\begin{figure}
  \centering 
  \includegraphics[width=0.95\columnwidth]{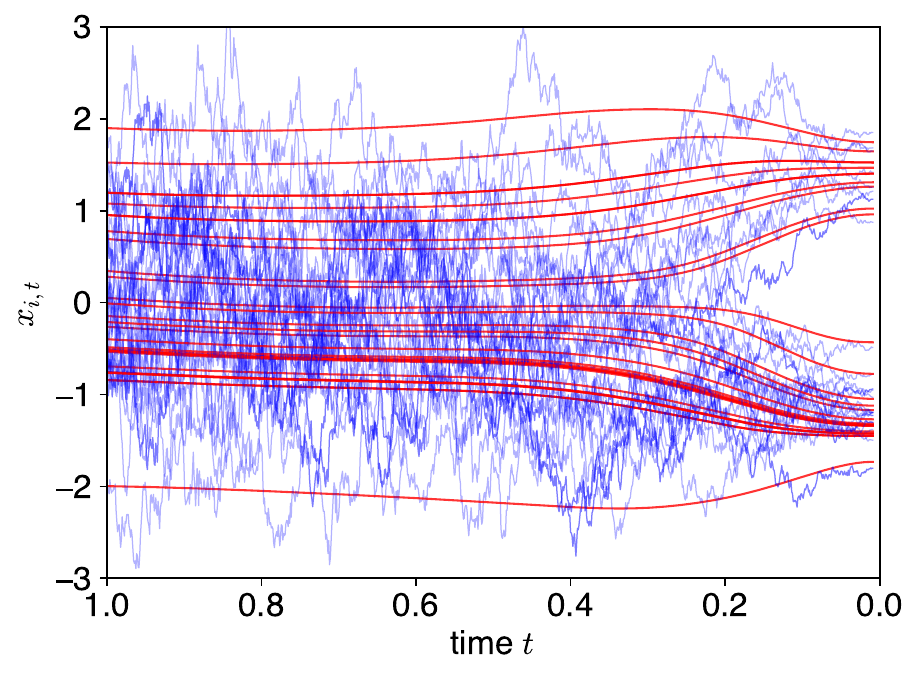}
  \caption{Deterministic trajectories of the solutions of the
    ODE~\eqref{eq:ODE} with the same initial values as chosen in
    Fig.~\ref{fig:reverse}.}
  \label{fig:reverse_ode}
\end{figure}

Now, we are ready to alter the SDE with a free parameter.  The
Fokker-Planck equation corresponding to the SDE~\eqref{eq:SDErev} is
immediately deduced from the above-mentioned derivation of the
Fokker-Planck equation as
\begin{equation}
  \dot{p}_t(\bx) = -\bnabla\cdot \biggl[ \Bigl( \bfunc(\bx,t)
  - g(t)^2 \bs_t(\bx) \Bigr) p_t(\bx) + \frac{g(t)^2}{2} \bnabla p_t(\bx)
  \biggr]\,.
\end{equation}
Here, we dropped $\btheta$ to simplify the notation.  We can split the
last term as
\begin{equation}
  \frac{g(t)^2}{2}\bnabla p_t(\bx)
  = \frac{\lambda^2 g(t)^2}{2}\bnabla p_t(\bx)
  + \frac{1-\lambda^2}{2} g(t)^2 \bnabla p_t(\bx)\,.
\end{equation}
The latter term is further rewritten as
\begin{equation}
  \frac{1-\lambda^2}{2} g(t)^2 \bnabla p_t(\bx) =
  \frac{1-\lambda^2}{2} g(t)^2 (\bnabla \ln p_t(\bx))\, p_t(\bx)\,.
\end{equation}
Because this final form is proportional to $p_t(\bx)$, we can regard
it as a part of the term involving
$\bfunc(\bx,t) - g(t)^2\,\bs_t(\bx)$.  Then, we can read the
corresponding SDE back from the rewritten Fokker-Planck equation,
yielding
\begin{equation}
  \dot{\bx}_t = \bfunc(\bx_t, t) - g(t)^2 \biggl( \bs_t(\bx_t)
  - \frac{1-\lambda^2}{2}\bnabla \ln p_t(\bx_t) \biggr)
  + \lambda\, g(t)\, \bar{\bxi}_t\,.
\end{equation}
If the learning is ideally perfect to realize
$\bs_t(\bx)=\bnabla \ln p_t(\bx)$, we can replace
$\bnabla \ln p_t(\bx_t)$ in the above SDE with $\bs_t(\bx_t)$.  After all, with this replacement, we arrive at the following class of SDEs with a free parameter $\lambda$:
\begin{equation}
  \dot{\bx}_t = \bfunc(\bx_t, t) - \frac{1+\lambda^2}{2} g(t)^2
  \bs_t(\bx_t; \btheta)  + \lambda\, g(t)\, \bar{\bxi}_t\,. 
\end{equation} 
Interestingly, if we choose $\lambda=0$, then the stochastic noise is
completely removed from the differential equation; that is, we can
perform the sampling procedure using the deterministic
equation~\cite{song2020denoising}:
\begin{equation}
  \dot{\bx}_t = \bfunc(\bx_t, t) - \frac{1}{2} g(t)^2 \bs_t(\bx_t; \btheta)\,.
  \label{eq:ODE}
\end{equation}
Without noises, the solutions of the ODE behave smoothly, as shown by
the red lines in Fig.~\ref{fig:reverse_ode}.  We note that the red
lines start with the initial values, $\{x_{i,T}\}$, chosen to be the
same as in Fig.~\ref{fig:reverse}.

% --- figure ---%
% --- figure ---%

Now, it is important to note that the time-dependence in
$\bs_t(\bx_t)$ should be properly adjusted within the finite time
interval $(0,T]$.  For example, if we continue solving
Eq.~\eqref{eq:ODE} until equilibration is achieved, then
$\dot{\bx}_t=0$ at $t\to-\infty$ leads to the condition,
$\bs_t(\bx_t)=2\bfunc(\bx_t)/g(t)^2|_{t\to-\infty}
=\bx_t/\sigma^2$.  Hence, all of $\{x_{i,t}\}$ eventually converges to
discrete points where the condition is satisfied.  Also, if we skip
the training procedure and simply plug $\bs_0(\bx)$ in
Eq.~\eqref{eq:ODE}, the solution of Eq.~\eqref{eq:ODE} does not follow
$p_0(\bx)$.  At the same time, these features imply that the
convergence of the ODE solution to follow $p_0(\bx)$ could be even more improved
with refinement of $\bfunc(\bx_t,t)$ and $g(t)$, though we do not
discuss this possibility in this review.

%%%%%%%%%%
\section{A Review of Stochastic Quantization}

In physics, among many, a direct analogue of the framework of the
diffusion model is found in the theory of stochastic quantization.  
Let us suppose that we have a $0$-dimensional field-theoretical model,
that is, a partition function given by a one-dimensional integral.  To
make a connection to the previous section about the DDPMs based on
Eq.~\eqref{eq:p0}, we shall take the following model:
\begin{equation}
  Z(a,b) = \int \rmd\phi\, \rme^{-S(\phi;a,b)}\,,\qquad
  S(\phi;a,b) = a\phi^2 + b\phi^4\,.
\end{equation}
Then, as a matter of fact, we are going to solve the same problem as
in the previous section using a different language and convention.
Here, $\phi$ has no coordinate dependence for simplicity, but the
generalization to a more realistic physical model is straightforward.  This model is useful as a
prototype of the quantum field-theoretical problem.  If $\phi$ has
only $t$ dependence, then the $(0+1)$ dimensional model can be
regarded as the quantum mechanical system.

For $\RE a<0$ which is physically more interesting than $\RE a>0$, the
analytical expression of the partition function is slightly changed
from Eq.~\eqref{eq:Zsol}.   We can compute the nonzero expectation
values of physical observables as
\begin{equation}
  \langle \phi^{2n}\rangle_{\mathrm{c}}
  = (-1)^n \frac{\partial^n}{\partial a^n} \ln Z(a,b)\,.
\end{equation}
In particular, for $(a,b)=(-4,1)$, we can find the analytical
expression for $\langle\phi^2\rangle_{\mathrm{c}}$:
\begin{equation}
  \langle\phi^2\rangle_{\mathrm{c}}
  = \frac{9}{8} + \frac{I_{3/4}(z) + I_{5/4}(z)
  +\tfrac{1}{8} I_{1/4}(z) - \tfrac{1}{8} I_{-1/4}(z)}
  {I_{1/4}(z) + I_{-1/4}(z)}
  \,.
\end{equation}
We note that this setup with $\RE a<0$ corresponds to a physical
system with spontaneous symmetry breaking.  Precisely speaking, there
is no spontaneous symmetry breaking unless the degrees of freedom are
infinitely large, and yet, we can see a bifurcation in numerical
simulations.

Since the problem is as elementary as one-dimensional integral, the
numerical integration is not difficult at all.  This situation will be
totally changed once the sign problem occurs as we will address
later.  For our present analysis with $(a,b)=(-4,1)$, we can easily
figure out the first two expectation values, for example, as
\begin{equation}
  \langle \phi^2\rangle_{\mathrm{c}} \approx 1.83534\,,\qquad
  \langle \phi^4\rangle_{\mathrm{c}} \approx 0.552211\,.
  \label{eq:exact}
\end{equation}
We could continue the calculations to higher powers if necessary, but
for the purpose of benchmark test, these first two expectation values
should suffice.

Now, let us explain how the stochastic quantization works.  In the
stochastic quantization, instead of performing the functional
integral, the key ingredient is the SDE along the fictitious time (or
the quantum axis), which is denoted by $\tau$ here.  The stochastic field
variable is defined by the solution of the SDE as
\begin{equation}
  \dot{\phi}_\eta(x,\tau)
  = -\frac{\delta S[\phi_\eta]}{\delta\phi_\eta(x,\tau)}
  + \eta(x,\tau)\,.
  \label{eq:sq_SDE}
\end{equation}
Generally, the field variables are functions of spacetime, $x$,
including both spatial and temporal coordinates.  Although our toy
model has no $x$ dependence, in the review part, we shall keep the
general notation with $d$-dimensional spacetime.  Here, $\eta(x,\tau)$
is the stochastic noise which satisfies,
\begin{equation}
  \begin{split}
    & \langle \eta(x,\tau)\rangle_\eta = 0\,,\\
    & \langle \eta(x_1,\tau_1)\eta(x_2,\tau_2)\rangle_\eta
    = 2\delta^{(d)}(x_1-x_2) \delta(\tau_1-\tau_2)\,,
  \end{split}
\end{equation}
where $\langle\cdots\rangle_\eta$ denotes the average over the noise
$\eta(x,\tau)$.  In the path-integral formalism, this can be
represented as
\begin{equation}
  \langle A[\eta] \rangle_\eta :=
  \frac{\displaystyle \int \calD\eta\, A[\eta]\, 
    \exp\biggl[ -\frac{1}{4}\int \rmd^dx\,\rmd\tau\, \eta(x,\tau)^2 
    \biggr]}
  {\displaystyle \int \calD\eta\,
    \exp\biggl[ -\frac{1}{4}\int \rmd^dx\,\rmd\tau\, \eta(x,\tau)^2 
    \biggr]} \,.
  \label{eq:eta_integ}
\end{equation}
Then, the quantum expectation value is obtained from the noise
expectation value at infinitely large $\tau$:
\begin{equation}
  \langle\phi(x_1)\cdots \phi(x_k)\rangle
  = \lim_{\tau\to\infty} \langle\phi_\eta(x_1,\tau)\cdots
  \phi_\eta(x_k,\tau)\rangle_\eta\,.
\end{equation}
This is the calculation scheme in the stochastic quantization.

Alternatively, we can introduce the probability distribution from
\begin{equation}
  \langle\phi_\eta(x_1,\tau)\cdots\phi_\eta(x_k,\tau)\rangle_\eta
  =: \int \calD\phi\, p_\tau(\phi)\, \phi(x_1)\cdots \phi(x_k)\,.
\end{equation}
We are now sufficiently experienced to write down the corresponding
Fokker-Planck equation immediately.  That is, the identification of
$\bfunc = -\bnabla_\phi S[\phi]$ and $g(t)=\sqrt{2}$ in
Eq.~\eqref{eq:FP} leads to
\begin{equation}
  \dot{p}_\tau = \int \rmd^d x\, \frac{\delta}{\delta \phi(x,\tau)}
  \biggl( \frac{\delta S}{\delta\phi(x,\tau)}
  + \frac{\delta}{\delta\phi(x,\tau)} \biggr)\, p_\tau[\phi]\,.
\end{equation}
We can introduce a trick here to deform this evolution equation with a
free parameter.  Then, the SDE is also modified as
\begin{equation}
  \dot{\phi}_\eta(x,\tau) = -\frac{\delta S[\phi_\eta]}{\delta\phi_\eta(x,\tau)}
  - (1-\lambda^2) \frac{\delta}{\delta\phi_\eta(x,\tau)} \ln p_\tau(\phi_\eta)
  + \lambda \eta(x,\tau)\,.
\end{equation}
Here again, we can take the limit of $\lambda=0$, so that the
$\tau$-evolution should become deterministic.  That is,
\begin{equation}
  \dot{\phi}(x,\tau) = -\frac{\delta S[\phi]}{\delta \phi(x,\tau)}
  - \frac{\delta}{\delta\phi(x,\tau)} \ln p_\tau[\phi] \,.
  \label{eq:m_ODE}
\end{equation}
Obviously, the $\tau$-evolution of $\phi(x,\tau)$ ceases when
$p_{\tau}[\phi]\approx \rme^{-S[\phi]}$.  So, once $p_\tau[\phi]$ or
the derivative of $\ln p_\tau[\phi]$ (which is a counterpart of the
score function) is trained, the stochastic nature can be completely
eliminated from the formalism.  Here again, we emphasize that the
quantum nature is captured by the entire $\tau$-evolution of
$p_{\tau}[\phi]$.  In this review, we will not pursue this issue, and
the interested readers can consult the recent work; see especially
Sec.\ 3.3 in the literature~\cite{Wang:2023exq} for the effective
action from the probability flow ODE formulation.  Although
Eq.~\eqref{eq:m_ODE} is not well known in physics, it deserves
extensive investigations.

\begin{figure}
  \centering 
  \includegraphics[width=0.8\columnwidth]{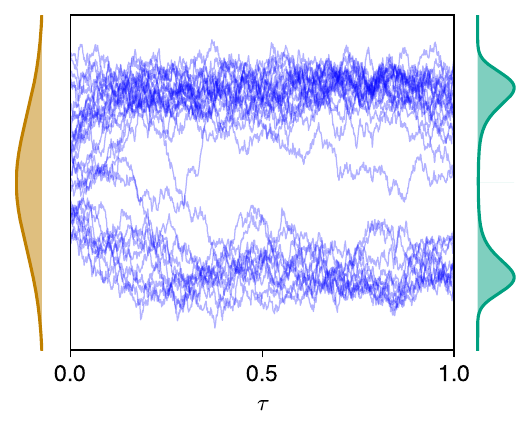}
  \caption{Randomly sampled 30 trajectories in the stochastic 
    quantization with the initial condition given by the normal 
    distribution.}
  \label{fig:traj_norm}
\end{figure}

For the practical application, we need the initial condition.  In the
literature, the null initial condition is the conventional choice, i.e.,
\begin{equation}
  p_{\tau=0}[\phi] = \prod_x\, \delta[\phi(x)]\,,
\end{equation}
but this form is not mandatory.  Actually, the above choice of the
singular form is numerically inconvenient and we can better start with
a more regular form such as the normal distribution.  Then,
Fig.~\ref{fig:traj_norm} shows the randomly sampled trajectories when
we choose $\Delta \tau=T/N_{\mathrm{step}}$ with
$N_{\mathrm{step}}=2^{10}$.  We immediately realize that the
convergence from the normal distribution to the original distribution,
$p_0(x)$, is qualitatively similar to the behavior we have seen in
Fig.~\ref{fig:reverse}.  The quantitative differences are caused by
the absence of $\bfunc(\bx,t)$ and $g(t)$ in the standard formulation
of the stochastic quantization.

% --- figure ---%
% --- figure ---%

It should be noted that a technique similar to introducing $g(t)$ is
known in the context of the stochastic quantization.  The noise
fluctuation in Eq.~\eqref{eq:eta_integ} could be regularized by a
kernel as follows,
\begin{align}
  & \int\calD\eta\, A[\eta]\, \exp\biggl[-\frac{1}{4}\int \rmd^d x\, 
  \eta(x,\tau)^2 \biggr] \notag\\
  & \to \int\calD\eta_K\, A[\eta_K]\, \exp\biggl[-\frac{1}{4}\int \rmd^d x\, 
  \rmd^d y\, \eta_K(x,\tau) K(x,y) \eta_K(y,\tau)\biggr]\,.
\end{align}
Equivalently, $\eta(x,\tau)$ itself is kept as the Gaussian noise and
the (square-root of) kernel $K(x,y)$ could be multiplied to the SDE in
Eq.~\eqref{eq:sq_SDE}.  If we use this modified noise, the SDE should
be altered as
\begin{equation}
  \dot{\phi}_{\eta_K}(x,\tau) = -\int \rmd^d y\,
  K(x,y)\, \frac{\delta S[\phi_{\eta_K}]}{\delta\phi_{\eta_K}(x,\tau)}
  + \eta_K(x,\tau)\,.
  \label{eq:sq_SDE2}
\end{equation}
This modification is sometimes required;  the na\"{i}ve application of
the method to fermions breaks down and it is convenient to
\textit{bosonize} the SDE with an appropriate kernel.

\begin{figure}
  \centering 
  \includegraphics[width=0.9\columnwidth]{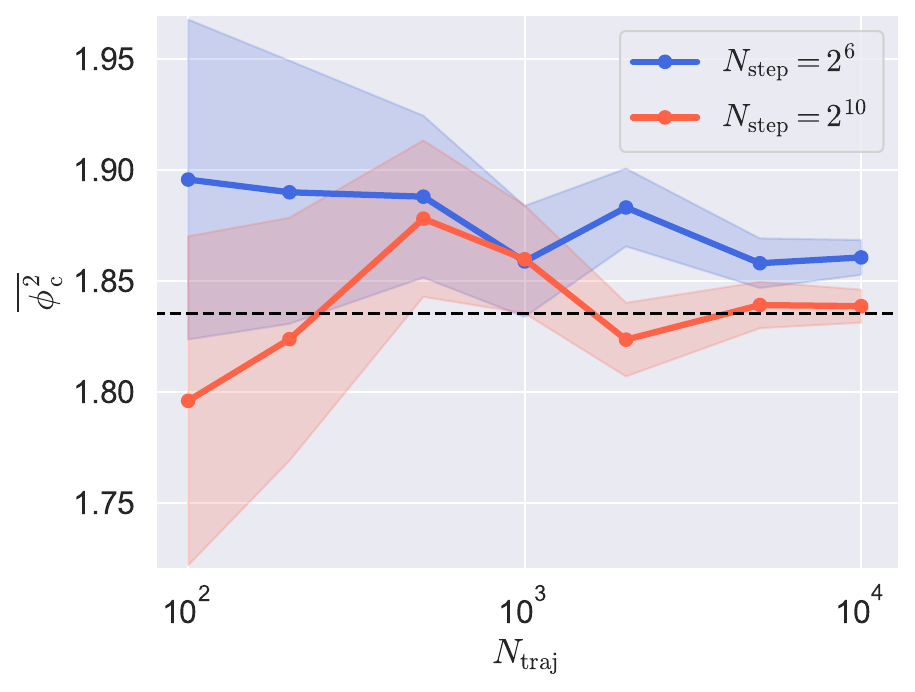}
  \caption{Ensemble average of $\phi^2$ over
    randomly sampled $N_{\mathrm{traj}}$ trajectories with different
    time steps.  The $1\sigma$ band is estimated from
    $N_{\mathrm{traj}}$ trajectories.  The dashed black line
    represents the exact answer.}
  \label{fig:conv2}
\end{figure}

\begin{figure}
  \centering 
  \includegraphics[width=0.9\columnwidth]{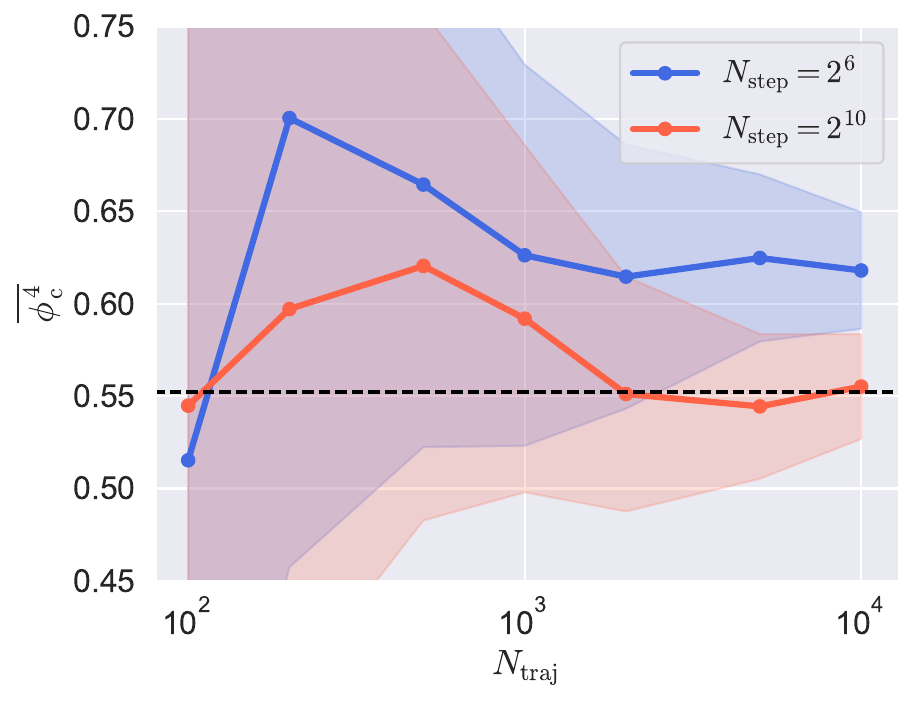}
  \caption{Ensemble average of the connected part of $\phi^4$ over
    randomly sampled $N_{\mathrm{traj}}$ trajectories with different
    time steps.  The $1\sigma$ band is estimated from
    $N_{\mathrm{traj}}$ trajectories.  The dashed black line
    represents the exact answer.}
  \label{fig:conv4}
\end{figure}

Let us turn back to the discussions about Fig.~\ref{fig:traj_norm}.
From the final distribution of the stochastic trajectories, we can
compute the average values of $\phi^{2n}$.  Here, we specifically
evaluate:
\begin{align}
  \overline{\phi^2}_{\mathrm{c}}
  &:= \frac{1}{N_{\mathrm{traj}}} \sum_{i=1}^{N_{\mathrm{traj}}}
    \phi_{i, T}^2 \,,\\
  \overline{\phi^4}_{\mathrm{c}}
  &:= \frac{1}{N_{\mathrm{traj}}} \sum_{i=1}^{N_{\mathrm{traj}}}
    \phi_{i,T}^4  -  \Bigl(\overline{\phi^2}_{\mathrm{c}}\Bigr)^2\,,
\end{align}
using randomly sampled $N_{\mathrm{traj}}$ trajectories.  We then
compare $\overline{\phi^2}_{\mathrm{c}}$ and
$\overline{\phi^4}_{\mathrm{c}}$ with the exact values in
Eq.~\eqref{eq:exact}.  We expect that these average values should
converge to the exact answer as $N_{\mathrm{traj}}$ increases.  We
quantify this tendency by making plots of
$\overline{\phi^{2n}}_{\mathrm{c}}$ as functions of
$N_{\mathrm{traj}}$ as shown in Figs.~\ref{fig:conv2} and
\ref{fig:conv4}.

% --- figure ---%
% --- figure ---%

% --- figure ---%
% --- figure ---%

From these results in Figs.~\ref{fig:conv2} and \ref{fig:conv4}, as
expected, we can conclude that the average values over the
trajectories certainly approach the exact answers as
$N_{\mathrm{traj}}$ increases for sufficiently large
$N_{\mathrm{step}}$.  Here, we fixed $T$ and decreased $\Delta \tau$
for larger $N_{\mathrm{step}}$.  We can alternatively increase $T$
with fixed $\Delta\tau$.

It is rather perplexing that the convergence appears not so fast;  even in
this simplest toy model, the average values with
$N_{\mathrm{step}}=2^{10}$ and $N_{\mathrm{traj}}=10^4$ are not
satisfactorily close to the exact answers.  To quantify this, we
estimated the error bar from
$\sigma=\sqrt{\overline{\phi^4}_{\mathrm{c}}/(N_{\mathrm{traj}}-1)}$
as displayed by the band in Fig.~\ref{fig:conv2}.  For $N_{\mathrm{step}}=2^{10}$
and $N_{\mathrm{traj}}=10^4$, we find that
$\overline{\phi^2}_{\mathrm{c}} = 1.8387 \pm 0.0075$, while the exact
answer is $\approx 1.83534$, which actually shows good agreement.  In the same
way, we estimated the error bar for $\overline{\phi^4}_{\mathrm{c}}$
to conclude $\overline{\phi^4}_{\mathrm{c}} = 0.5552\pm 0.0283$ for
the exact answer $\approx 0.55221$.  In this case, the numerical
result happens to be close to the exact answer, but the error bar is
still large and the numerical agreement seems to be
accidental.

\begin{figure}
  \centering 
  \includegraphics[width=0.9\columnwidth]{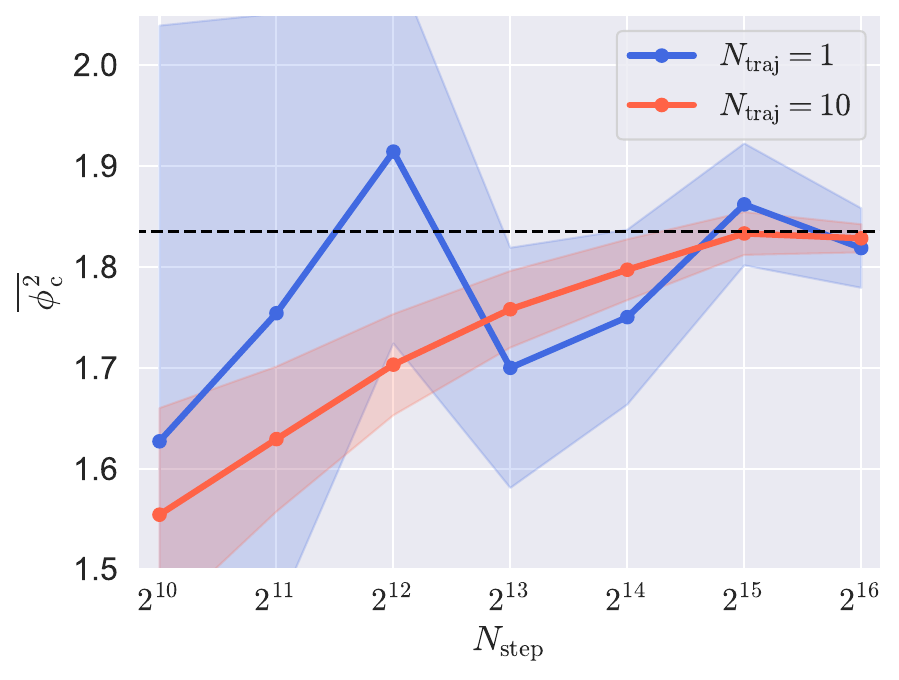}
  \caption{Time average of the connected part of $\phi^2$ over
    $N_{\mathrm{step}}$ time steps and $N_{\mathrm{traj}}$
  trajectories.   The $1\sigma$ band is estimated from 100 runs.  The
  dashed black line represents the exact answer.}
  \label{fig:convtime2}
\end{figure}

To improve the convergence problem, the common strategy is
to replace the \textit{ensemble average} over trajectories with the
\textit{time average} of stochastic evolution.  In principle, if the
time extent is large enough, even a single trajectory should reproduce
the correct answer in this way.  Let us check how this strategy works.  We make a
plot similar to Fig.~\ref{fig:conv2} to demonstrate the convergence
properties of the time-averaged value.  We calculated not only
$\overline{\phi^2}_{\mathrm{c}}$ but also
$\overline{\phi^4}_{\mathrm{c}}$, but it is sufficient to show the
comparison of $\overline{\phi^2}_{\mathrm{c}}$ for the present
demonstration.  Figure~\ref{fig:convtime2} presents the results for
$N_{\mathrm{traj}}=1$ and $N_{\mathrm{traj}}=10$ with the error bar
estimated from fluctuations with 100 independent runs.  From these
figures, we see that the time average of even the single trajectory
($N_{\mathrm{traj}}=1$) can approach the exact answer if
$N_{\mathrm{step}}$ is large enough.  Of course, the agreement with
the exact answer is more guaranteed by further taking the ensemble
average over $N_{\mathrm{traj}}>1$ trajectories.

% --- figure ---%
% --- figure ---%

%%%%%%%%%%
\section{Sign Problem}

The reinterpretation of the stochastic quantization and the diffusion
model has been considered~\cite{Wang:2023exq} which aims to improve
the efficiency to sample quantum lattice field configurations.  The
applications of two seemingly different but essentially equivalent
formulations have just started recently, and further developments
should await to be revealed.

Among various possibilities, one direction more exciting than
improving the efficiency is an attempt to evade the sign problem.  As
long as $\rme^{-S[\phi]}$ is positive definite, we can give it a meaning as
the probability, $p_\tau[\phi] \sim \rme^{-S[\phi]}$.  However, in
many interesting physical systems, $\rme^{-S[\phi]}$ is not
necessarily positive definite.  For example, the real-time evolution
requires the Minkowskian formulation in which the functional
integral involves $\rme^{\rmi S[\phi]}$.  Because of this complex
nature, $\rme^{\rmi S[\phi]}$ is an oscillatory function and the
Monte-Carlo integration algorithm breaks down.  Another example is
found in fermionic systems at finite chemical potential;  see
reviews~\cite{Muroya:2003qs,Fukushima:2010bq,Aarts:2015tyj,Nagata:2021ugx}.

Although the probability interpretation loses its meaning, the stochastic
quantization scheme still looks feasible.  The only extension is that
the trajectories may spread over the complex plane;  in other words,
$\phi_\eta$ may become a complex-valued function.  This generalized
stochastic quantization method is referred to as the Complex Langevin
Equation (CLE) approach.  Actually, the CLE application to the Strong
Interaction has a long history traced back to 1985 by a pioneering
work~\cite{Karsch:1985cb}.  Since then, the interest in nuclear
physics was revived around 2010; see, e.g., a discussion on the
prospect of gauge-cooling technique~\cite{Aarts:2013uxa}.  Continued
and latest applications include the quark-flavor number dependence of
finite-density matter~\cite{Namekawa:2021qtg} and full real-time
simulations of strongly interacting
system~\cite{Boguslavski:2023unu}.

The problem in the CLE method is that the results may not converge or even the converged results may not be correct.  The convergence
criterion has been known; see some
discussions~\cite{Aarts:2013uza,Shimasaki:2016ygt}.

\begin{figure}
  \centering 
  \includegraphics[width=0.85\columnwidth]{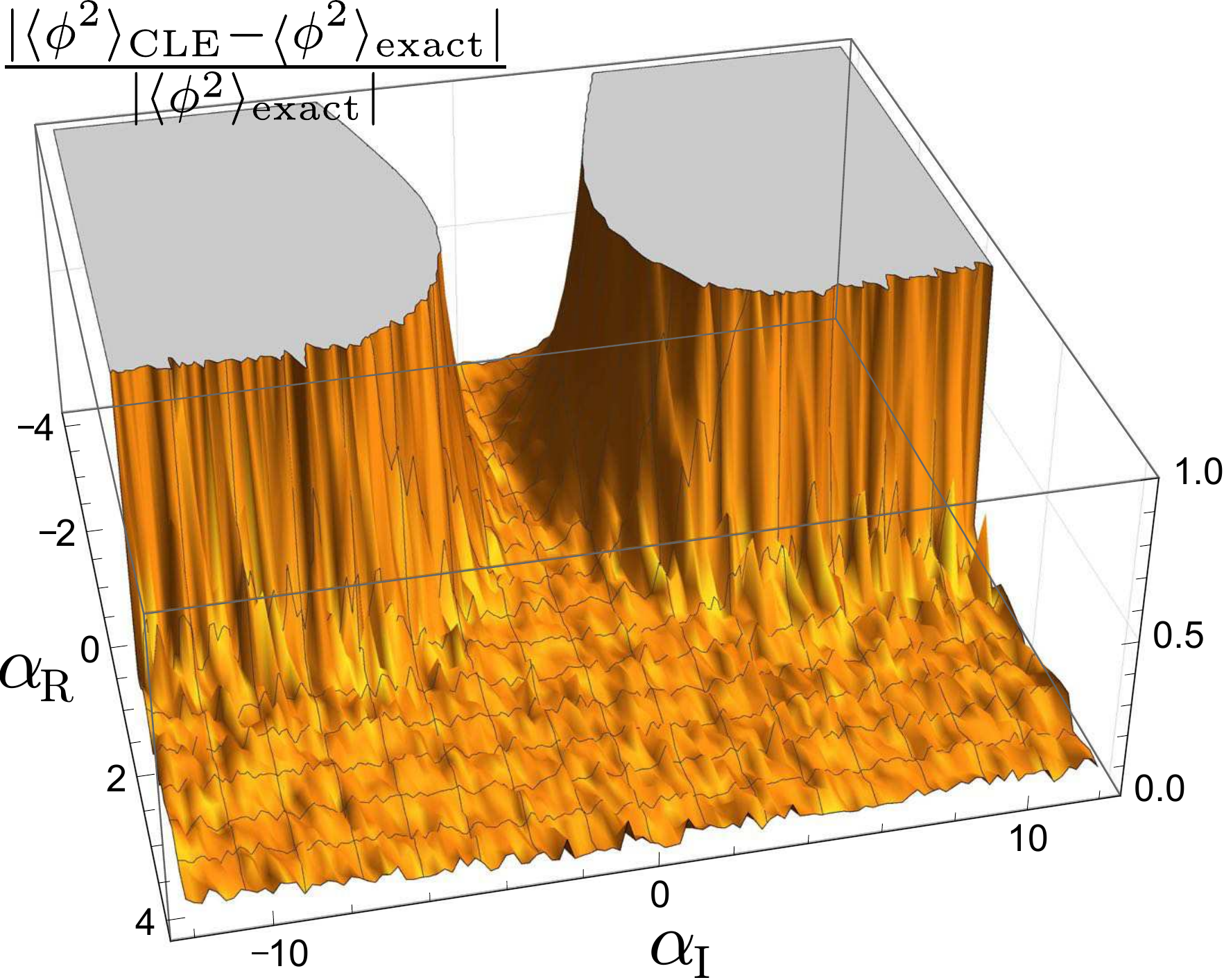}
  \caption{Comparison to the exact answer in complex parameter space.
    In the present notation, the quadratic coefficients are related as
    $\alpha_{\mathrm{R/I}}=a_{\mathrm{R/I}}/2$ and the quartic
    coefficient is fixed as $\beta=b/4=1$.
  Figure is adapted from the paper~\cite{Abe:2016hpd}.}
  \label{fig:cle}
\end{figure}

We can introduce the sign problem into our simple toy model.  It is an
intriguing question what would happen if we generalize the model
parameters on the complex plane.  Here, let us take $a=\ar + \rmi \ai$
with $\ar=-4$ fixed and we change $\ai$ to control the degree of
complexification.  Actually, this model has been carefully studied in
the paper~\cite{Abe:2016hpd} with a slightly different notation;
$\alpha=a/2$ and $\beta=b/4$ in the potential coefficients and
$\beta=1$ was chosen there~\cite{Abe:2016hpd}.  Then, the
breakdown of the CLE method has been quantified as shown in
Fig.~\ref{fig:cle} as a function of the real and imaginary parts of
$\alpha$.

% --- figure ---%
% --- figure ---%

As long as $\alpha_{\mathrm{R}} > 0$, the convergence property is good
and regardless of complexification with $\alpha_{\mathrm{I}}\neq 0$,
the CLE can converge to the correct answer of
$\langle\phi^2\rangle_{\mathrm{c}}$.  In the region with
$\alpha_{\mathrm{R}} < 0$, the sign problem occurs and the CLE breaks
down except for the region with
$|\alpha_{\mathrm{I}}| \ll |\alpha_{\mathrm{R}}|$ where the method is
reduced to the real stochastic quantization.  In this way, as
intuitively expected, we understand that the sign problem turns out to
be severe in the region with $\alpha_{\mathrm{R}} < 0$ and
$|\alpha_{\mathrm{I}}| > |\alpha_{\mathrm{R}}|$.  Strictly speaking,
this conclusion is valid for $\langle\phi^2\rangle_{\mathrm{c}}$ and
higher-order operator expectations might be more sensitive to the sign
problem.

% --- figure ---%
\begin{figure}
  \centering 
  \includegraphics[width=0.9\columnwidth]{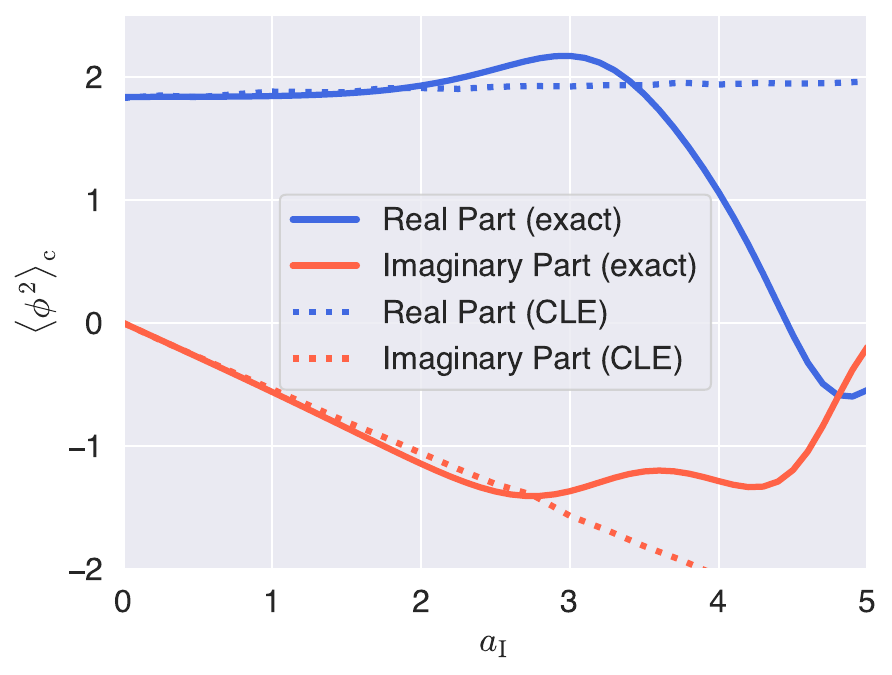}
  \caption{CLE simulations compared to the exact numerical values as a 
    function of $\ai$ with $\ar=-4$ and $b=1$ fixed.}
  \label{fig:complex_exact}
\end{figure}
% --- figure ---%

Now, let us carry out the direct CLE simulation using the present
setup of the toy model.  We fix $\ar=-4$ and $b=1$ (which corresponds
to $\alpha_{\mathrm{R}}=-8$ and $\beta=4$ in the previous
convention~\cite{Abe:2016hpd}), and calculate the real and imaginary
parts of $\langle\phi^2\rangle_{\mathrm{c}}$.  The results are
summarized in Fig.~\ref{fig:complex_exact}.  For this simulation, we
chose $N_{\mathrm{traj}}=10$ and $N_{\mathrm{step}}=2^{16}$ and took
the time-average.  Also, we used a discretization scheme with some
resummation to avoid run-away behavior of trajectories.  For technical
details, see relevant discussions in the anharmonic oscillator
simulation~\cite{Anzaki:2014hba}.

In view of Fig.~\ref{fig:complex_exact}, the sign problem certainly
gets worse for $\ai \gtrsim |\ar|$.  Of course, the best would be providing the
solution of the sign problem, but the sign problem is known to be
NP-hard~\cite{PhysRevLett.94.170201}, and simple solutions within
reasonable time scale would be unattainable.  Then, the second best
would be the identification of the difficulty in the formalism.  In
this context, it is shown that the analysis of the Lefschetz thimble
can make it clear where the CLE may fail~\cite{Hayata:2015lzj}.

The idea of complexifying the path integral using  the Lefschetz
thimble was demonstrated in the seminal work~\cite{Witten:2010zr},
which was successfully implemented for numerical simulations in
quantum field theories~\cite{Cristoforetti:2012su}.  Since this is
nothing but the higher-dimensional extension of the complex analysis
for the one-dimensional integral, we do not have to consider thimble
structures for the present setup of the one-dimensional problem.  Let
us introduce a complex variable; $z=\RE\phi + \rmi \IM\phi$.  We can
deal with the complexified action, $S[\phi]\to S[z]$, then.  We should
find the critical points, $z_i$, by solving the saddle-point
condition,
\begin{equation}
  S'[z] \bigr|_{z=z_\sigma} = 0\,.
\end{equation}
In our present case with $S[z]=az^2 + bz^4$, we should get three
critical points at $z=0$, $z=\pm\sqrt{-a/(2b)}$.  The steepest descent
cycles are defined with a time-like variable, $\tau$, as
\begin{equation}
  \calI_\sigma := \biggl\{\; z(\tau) ~ \Big| ~
  \frac{\rmd z}{\rmd \tau} = \overline{\frac{\partial S}{\partial
      z}}\,,\; z(\tau\to-\infty) = z_\sigma\biggr\}\,.
  \label{eq:thimble}
\end{equation}
The nicest feature about the steepest descent cycles is that the phase
oscillation is suppressed on it, which is almost obvious from
$(\rmd/\rmd\tau) \IM S \propto (\rmd/\rmd\tau)(S-\overline{S})
=S'(\rmd z/\rmd\tau)-\overline{S}'(\rmd \overline{z}/\rmd\tau)
=S'\overline{S'} - \overline{S}'S'=0$\,.  The original integral can be
safely deformed to the sum of integrals on the steepest descent cycles
or the Lefschetz thimbles.  Once this rewriting is complete, it is
only the residual sign problem which causes obstacle in the numerical
simulation~\cite{Fujii:2013sra}.

% --- figure ---%
\begin{figure}
  \centering 
  \includegraphics[width=0.7\columnwidth]{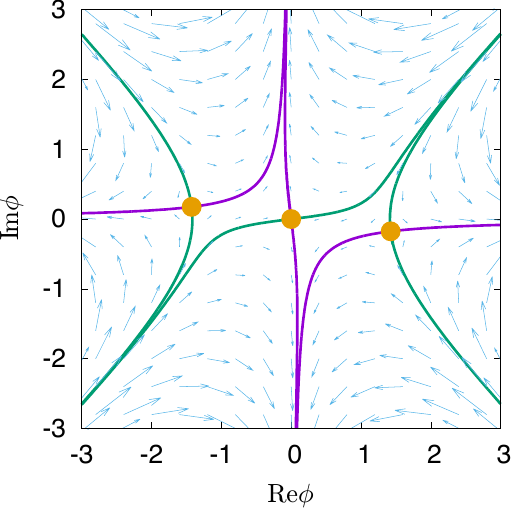}
  \caption{Structure of the steepest descents (purple) and the
    steepest ascents (green) with the saddle points (orange dots) 
    for $\ai=1$ with $\ar=-4$ and $b=1$ fixed.  The slope,
    $\overline{S'}$, is indicated by the arrows.}
  \label{fig:thim1}
\end{figure}
% --- figure ---%

% --- figure ---%
\begin{figure}
  \centering 
  \includegraphics[width=0.7\columnwidth]{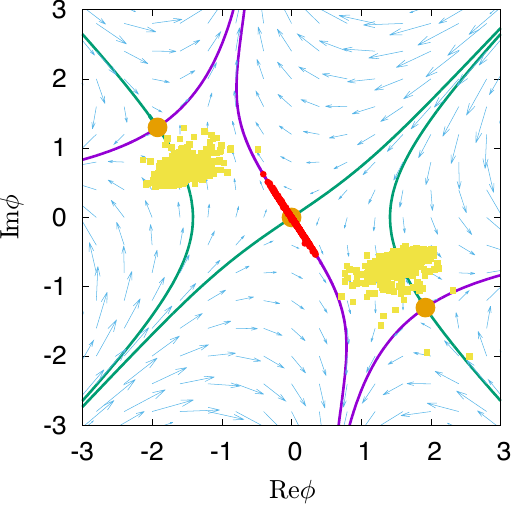}
  \caption{Same as Fig.~\ref{fig:thim1} for $\ai=10$ with $\ar=-4$ and
    $b=1$ fixed.  Scattering data represent the CLE sampled points
    without the modified kernel (yellow) and with the modified kernel (red).}
  \label{fig:thim10}
\end{figure}
% --- figure ---%

Figures~\ref{fig:thim1} and \ref{fig:thim10} show the structures of
the steepest descents by the purple lines and the steepest ascents by
the green lines as well as the slope, $\overline{S'}$, indicated by
small arrows (the length of the arrows is square-root proportional to
the modulus of the slope).  When the sign problem is minor with
$\ai=1$, three critical points are almost along the real axis as seen
in Fig.~\ref{fig:thim1}.  The original integration path should be
deformed into three pieces attached to three critical points.  For the
calculation of $\langle\phi^2\rangle_{\mathrm{c}}$, the contribution
from $z=0$ is suppressed, and the saddle-point approximation at
$z=\pm\sqrt{-a/(2b)}$ leads to
\begin{equation}
  \langle\phi^2\rangle_{\mathrm{c}} \simeq
  \frac{-(a/b) \rme^{a^2/(4b^2)}}{1+2\rme^{a^2/(4b^2)}}\,.
  \label{eq:saddle}
\end{equation}
If $|2\rme^{a^2/(4b^2)}| \gg 1$, then we can further simplify the
above estimate as
$\langle\phi^2\rangle_{\mathrm{c}}\simeq -a/2=2-\rmi \ai/2$.  This
formula nicely explains the small-$\ai$ behavior of
Fig.~\ref{fig:complex_exact}.

The CLE method clearly breaks down for $\ai\gtrsim 4$ as quantified in
Fig.~\ref{fig:complex_exact}, and the configuration of the steepest
descents/ascents for $\ai=10$ is shown in Fig.~\ref{fig:thim10}.  At
first glance, we see no qualitative difference from
Fig.~\ref{fig:thim1}.  The saddle point approximation for $\ai=4$ gives
$\RE\langle\phi^2\rangle_{\mathrm{c}}\approx 0.78$ and 
$\IM\langle\phi^2\rangle_{\mathrm{c}}\approx -2.57$.  So, the
saddle-point approximation is not good any more.  As a matter of fact,
the breakdown of the CLE method and the deviation from the leading-order
saddle-point approximation are somehow related.  The weight of the
saddle-point contributions at $z=\pm\sqrt{-a/(2b)}$ is
$|\rme^{a^2/(4b^2)}|=\rme^{4-\ai^2/4}$, so that the exponent changes
its sign at $|\ai|=4$.  Thus, around $\ai\sim 4$, the saddle-point
approximation is no longer effective with such a small exponent, and the relative
weight between $1$ from $z=0$ and $\rme^{a^2/(4b^2)}$ from
$z=\pm\sqrt{-a/(2b)}$ is flipped.

% --- figure ---%
\begin{figure}
  \centering 
  \includegraphics[width=0.85\columnwidth]{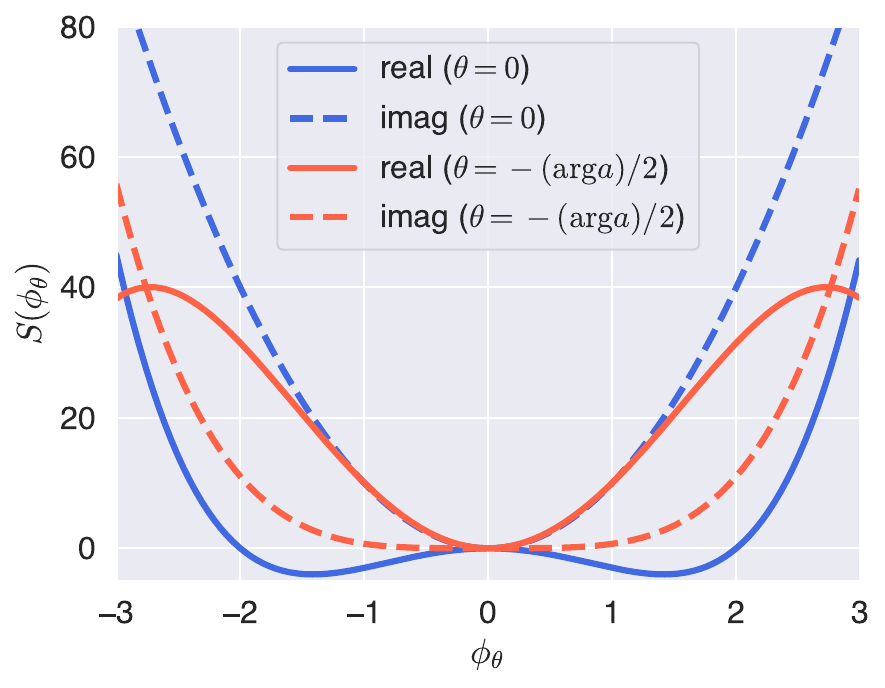}
  \caption{Action or potential energy for $\ai=10$ along $\IM{\phi}=0$, 
    i.e., $\theta=0$ indicated by the blue solid (real part) and dashed 
    (imaginary part) lines.  With the kernel, the red solid (real 
    part) and dashed (imaginary part) lines represent the action along 
  $\arg(a\phi^2)=0$, i.e., $\theta=-(\arg a) /2$.}
  \label{fig:action}
\end{figure}
% --- figure ---%

In view of Fig.~\ref{fig:complex_exact}, the CLE results can be
understood if $1$ in the denominator in Eq.~\eqref{eq:saddle} is
dropped, though it should not be dropped in reality.  Indeed, this observation can be confirmed by the
distribution of the CLE sampled points overlaid by yellow dots on
Fig.~\ref{fig:thim10}.  It is obvious that the CLE method fails to
collect the contribution near $z=0$.  A more comprehensive analysis is
found in the literature~\cite{Nagata:2016vkn}.  If we make a similar
scattering plot on Fig.~\ref{fig:thim1}, it is also the case that the
$z=0$ point repels sampled data.  Of course, such repulsion around
$z=0$ is perfectly reasonable for $\ai\approx 0$, for $\rme^{-S[z]}$
has maxima at $z=\pm\sqrt{-a/(2b)}$, while the $z=0$ point corresponds
to the minimum, which is least favored naturally.  In terms of the
Lefschetz thimble, the repulsion simply means a small relative
weight.  As $\ai$ grows up, the saddle points are aligned along the
path with the phase angle by $-(\arg a)/2$.  It has been shown that the
CLE method with the rotated variable from $\phi$ to $\phi_\theta$ by
$-(\arg a)/2$, that is,
\begin{equation}
  \phi =: \rme^{\rmi \theta} \phi_\theta\,, \qquad
  \theta = -(\arg a)/2\,,
  \label{eq:v_change}
\end{equation}
can improve the convergence to the correct answer~\cite{Abe:2016hpd}.
Then, we see $a\phi^2 = |a|\phi_\theta^2$.  In fact, in the vicinity
of $z=0$, we draw the action $S[z]$ (or it could be called the
potential) as a function of this rotated variable $\phi_\theta$ in
Fig.~\ref{fig:action}.  Along this direction of $\phi_\theta$ with
$\theta=-(\arg a)/2$, the imaginary part of the action is flat and the
real part shows a minimum.  This makes a sharp contrast to the $\theta=0$ case that the real part has a double-well shape with a maximum at $\phi=0$.  Accordingly, for the case with $\theta=-(\arg a)/2$, the sampled points are certainly
localized around $z\simeq 0$ as seen from the red dots in
Fig.~\ref{fig:thim10}.  The important point is that the
SDE~\eqref{eq:sq_SDE} is not invariant under the variable change like
Eq.~\eqref{eq:v_change}.  In fact, as we already mentioned, the SDE
could be modified as Eq.~\eqref{eq:sq_SDE2} with a kernel and the
phase transformation in Eq.~\eqref{eq:v_change} corresponds to the
choice of the kernel as $K(x,y)= \rme^{2\rmi\theta}\delta(x-y)$.
Surprisingly, the choice of the variable or the kernel would affect
the final numerical output.  In the present case with $\ai=10$, for example,
the correct answer is
$\langle \phi^2\rangle \approx -0.01051 - 0.04931 \rmi$.  The
na\"{i}ve application of the CLE method gives a totally wrong result,
$\langle\phi^2\rangle_{\mathrm{CLE}} \approx
(1.98058\pm 0.02333) + (-5.06101\pm 0.00082)\rmi$, while
the modified CLE method with the kernel results in a much better value,
$\langle\phi_\theta^2\rangle_{\mathrm{CLE}} \approx
(-0.01901\pm 0.00079) + (-0.04503\pm 0.00184)\rmi$.

%%%%%%%%%%
\section{Speculative Prospects}

Now, we have seen a quite suggestive analogy between the stochastic
quantization and the diffusion model.  In the most direct application
of the analogy, the diffusion models can be utilized to generate the lattice
configurations efficiently.  Here, we have put more emphasis on the dark
side of the numerical simulation, namely, the sign problem.  Actually,
for the diffusion models too, the difficulty encountered with complex
variables is not necessarily an academic exercise.  Nowadays,
``quantum'' is such a fashionable keyword, and the crucial difference
between quantum and classical lies in the information of the complex
phase from where the interference effect emerges.  Therefore, if one
would like to \textit{quantumize} the generative models, the first
step would be to complexify the formalism.  Although the
complexification may not cause troubles in some parameter space, the
empirical rule we know from physics realms is that the
problem gets harder in the regimes with more interesting
contents.  We speculate that our knowledge about the CLE method should be
useful for future studies along these lines.  In particular, we have
established good understanding of the breakdown of the method in terms
of the Lefschetz thimble and demonstrated a potential resolution by means of the optimized kernel.

A more ambitious direction of research is to solve or tame the sign
problem by means of the machine-learning techniques.  There are some
preceding works to optimize the integration
path~\cite{Mori:2017pne,Mori:2017nwj}, for example.  The analogy to
the diffusion models may pave a novel passage to tackle the sign
problem.  In the CLE, the convergence problem occurs due to run-away
trajectories.  In the present simulation, we used a half-implicit
method to regularize such trajectories, but in general, there are always
unstable directions in complex plane.  This problem may be cured by
the ODE-type evolution without any fluctuations.  In gauge theories,
the gauge cooling is a technology to evade the run-away trajectories,
and if the ODE reformulation turns out to be effective, the algorithm
may be improved.  An even more radical speculation is the possibility
of accessing the analytical structures of the trained score function.
The score function provides us with a mapping between the normal
distribution and the probability distribution of our interest, which
is nothing but the procedure to \textit{solve} the theory in physics.
The mapping is also translated into the change from the original
variables to the optimal variables in the integral, and if the
optimization imposes the condition to suppress the phase oscillation
in a way as described below Eq.~\eqref{eq:thimble}, the flow to
de-complexify the theory may naturally find the paths along the
Lefschetz thimbles.  Alternatively, the optimized kernel could be found in the machine-learning assisted algorithm.

What is speculated here may sound too optimistic, but anything is not
mature yet.  There are many fascinating attempts in physics, and the
Lefschetz thimble is just one of them.  We did not mention here, but
the stochastic quantization has a useful mathematical structure of
supersymmetry (see discussions in the
literature~\cite{Fukushima:2015qza}).  We are making efforts to export
the physics wisdom to the diffusion models.  We would like to invite
interested physicists to this aspiring project and we hope that our
present review serves as an interpreter of two descriptions in physics and
machine learning.

\begin{acknowledgment}
  We would like to thank
  Yuji~Hirono,
  Jan~Pawlowski,
  Franz~Sattler,
  Akinori~Tanaka,
  Lingxiao~Wang,
  and Kai~Zhou
  for useful conversations.
  This work is partially supported by the JSPS KAKENHI Grant No.\ 22H05118.
\end{acknowledgment}

\profile{Syo Kamata}{is a project assistant professor at Department of
  Physics, The University of Tokyo.  The research subjects include the
Lefschetz thimble in low-dimensional quantum field theory, the exact
WKB, and the machine-learning analysis of the astrophysical observations.}
\profile{Kenji Fukushima}{is a professor at Department of Physics, The
  University of Tokyo.  The research subjects include the high-energy
  nuclear physics in extreme environments such as high temperature, high baryon density, strong
  magnetic field, etc.}

\bibliography{ddpm}
\bibliographystyle{jpsj}

\end{document}